# Non-covalent bonds between optical solitons in an optoacoustically mode-locked fiber laser: Analysis & modelling

W. He[1], M. Pang[1,*], D. H. Yeh[1], J. Huang[1], C. R. Menyuk[2], and P. St.J. Russell[1]

[1]*Max Planck Institute for the Science of Light, Staudtstrasse 2, 91058, Erlangen, Germany*

[2]*Department of Computer Science and Electrical Engineering, University of Maryland Baltimore County, Baltimore, Maryland 21250, USA*

*Corresponding author. E-mail: meng.pang@mpl.mpg.de

Binding of particles through weak (non-covalent) interactions plays a central role in multiple modern disciplines, from Cooper pairing of electrons in superconductivity to formation of supramolecules in biochemistry. Optical solitons, which are analogous in many ways to particles, arise from a balance between nonlinearity and dispersion and have only been known to be able to form molecule-like, localized structures through covalent interactions due to direct overlapping. Non-covalent binding of optical solitons that based on long-rang interactions have, however, so far escaped experimental observations. Long-range interactions between optical solitons have proven almost impossible to control, generally leading to disorder. In this paper, however we show that by tailoring weak, long-range soliton interactions, a large population of optical solitons in an optoacoustic mode-locking fiber laser can assemble into stable, highly-ordered structures that resemble biochemical supramolecules. We demonstrate here a theoretical model of the non-covalent binding mechanism, and provide some experimental results supporting this model.

## I. INTRODUCTION

Optical solitons, arising from a stable balance between nonlinear and dispersive effects, are analogous in many ways to particles. Pairs of solitons, propagating together in close proximity, can strongly interact through cross-phase modulation at their pulse tails, resulting in the formation of robust bound states that are frequently referred as soliton molecules or crystals[1-6] in analogy to chemical molecules formed by covalent bonds. The supramolecular assemblies of atoms or molecules which rely on weak, non-covalent bonds in the field of chemistry and biology have not yet, however, found their counterparts in terms of optical solitons. Although weak interactions between optical solitons have been studied in many systems[7-12], they have generally been regarded either as uncontrollable[8,13,14] or as sources of noise, leading to disordering of soliton sequences during propagation[9,15], consequently, the soliton supramolecules have so far escaped experimental observations. In this paper we report that in an optoacoustic mode-locking fiber laser, the long-range soliton interactions that arise from optoacoustic effects and dispersive wave perturbations can be carefully tailored, exerting attraction and repulsion forces between optical soliton in the laser cavity. The balance between these two forces leads to non-covalent bonds which enable the supramolecular assembly of a large number of optical solitons, resulting in complex supramolecular structures with soliton-soliton spacing hundreds of times longer than the individual soliton duration. In the experiments we also observe that the phased-locked soliton molecules, as fundamental elements, can also be incorporated into such supramolecular assembly, and the soliton supramolecules exhibit several key features[16,17], including structural flexibility, reversibility, elementary diversity and dynamic stability. The paper is organized as: we start from an analytical model that demonstrates the balanced long-range interactions between the solitons. Then, we present the some experiment results that support our theoretical model.

## II. NON-COVALENT BONDS BETWEEN SOLITONS: THEORY

### A. Optoacoustic mode-locking: the backbone of soliton supramolecules

We have previously demonstrated the concept of optoacoustic mode-locking using a solid-core photonic crystal fiber (PCF) inside a ring-cavity soliton fiber laser[18-21]. The acoustic resonance of the PCF-core, typically in few-GHz range, has effectively divided the relatively long fiber-cavity (FSR being a few MHz) into hundreds of identical time-slots, each trapping a single-soliton inside it, leading to equal soliton spacings as well as highly suppressed noise level[18,19]. Therefore, a stable, passive high-harmonic mode-locked laser has been obtained, settling the backbone of the soliton supramolecule. In the previous mode-locked lasers, only a single soliton can be trapped inside each time-slot, although some time-slots can be left blank without affecting the stable mode-locking state[21]. However, in a soliton supramolecule as will be presented in following sections, more than one solitons can be stably trapped within each time-slot (as sketched in FIG. 1 under the moving frame). Most significantly, these multiple solitons are bound by long-range, non-covalent interactions between them, leading to double-, triple- or even quadruple-soliton bound-state. In addition, throughout the entire round-trip time, different time-slots can trap different number of solitons, leading to macroscopic, complex soliton stream. The incremented degree-of-freedom in such optical structure originated from the interplay between different long-range interactions, which is described as follows.

**B. Non-covalent bonds between solitons: Basic mechanism**

The multi-soliton long-range bound-state existing within each time-slot relies on the non-covalent bonds that result from the balance between different long-range interactions. We consider the simplest case of such bound-state unit: the double-soliton unit, in which the two solitons are bound together with stabilized spacing between them. Such a bound-state is highlighted in the dash box in FIG. 1 using a moving frame. The 1st soliton, which is chosen as the "reference soliton" is located at a later time $t_1$ and the 2nd soliton is located at earlier time $t_2$. In one hand, the refractive index modulation induced by the acoustic wave leads to force of attractions between the two solitons (see FIG. 1). The effective force of attraction results in a slower group velocity of the 2nd soliton, which delays the 2nd soliton relative to the 1st one during propagation. On the other hand, the dispersive wave shed by the solitons during propagation, which is caused by the periodic perturbation in the laser cavity, would provide the "spring" effect (see FIG. 1), exerting a force of repulsion between the two solitons. At the position where the two countering forces balance out, a temporal trapping potential would appear for the 2nd soliton, and therefore a stable long-range bound-state of these two solitons can be realized (FIG. 1). In a same way, the second and third "springs" would appear in a cascaded way, leading to a triples- and even quadruple-soliton units (see FIG. 1), which have all been observed in experiments.

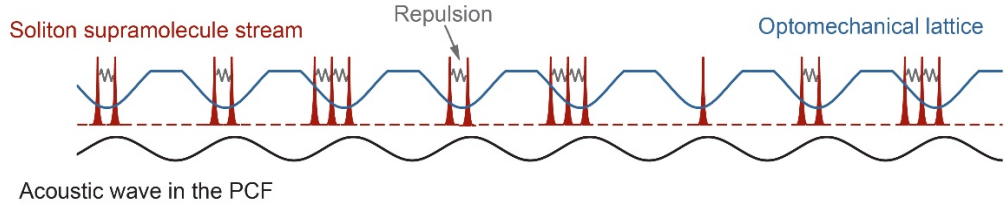

FIG. 1. Sketch of a typical soliton supramolecule stream under the moving frame. A long-lived acoustic wave co-propagate with the soliton stream, serving as an optomechanical lattice.

Since both type of interactions involved are long-lived (and thus long-range) in nature, the binding spacing can be much larger than the soliton duration, and moreover, no fixed phase-relation is required directly between the two solitons, such long-range binding resembles the non-covalent bonds in the fields of supramolecular biochemistry[16]. The simplest form of a soliton supramolecule stream is the all-double-soliton (ADS) case, in which all the time-slots are filled with identical double-soliton bound-state. In this case, the acoustic wave driven by the soliton stream would have a relatively simple form as we will see below.

**C. Force of attraction through acoustic wave modulation**

When the ADS supramolecule circulates in the laser cavity with a lattice frequency of $\Omega_a$, it coherently drives the mechanical core resonance in the PCF. The optically-driven mechanical vibration, in the form of a refractive index modulation, acts back on the driving solitons by varying their carrier frequencies[21]. In practice, we only consider the LP$_{01}$ optical mode and the R$_{01}$ mechanical resonance in the PCF core. The R$_{01}$ mechanical mode in the PCF has a resonant frequency of $\Omega_{01}$ and a mechanical bandwidth of $\Gamma_B$. In a reference frame moving with the soliton supramolecule, the mechanical vibration in the PCF core driven by the ADS supramolecule can be expressed as[21]:

$$b(t) = \frac{\gamma_e |Q| P_{av} \cos(\Omega_a \Delta t / 2)}{2\pi c n_0 A_{eff} \Omega_a \sqrt{4\delta_\Omega^2 + \Gamma_B^2}} \sin(\Omega_a t), \qquad (1)$$

where $b(t)$ is the material density variation as a function of time, $\gamma_e$ is the electrostrictive constant of silica, $c$ is the speed of light in vacuum, $n_0$ is the refractive index of silica, $A_{eff}$ is the effective mode area of LP$_{01}$ optical mode, $\delta$ is the frequency off-set ($\delta_\Omega = \Omega_a - \Omega_{01}$), and $P_{av}$ is the average optical power in the PCF. The overlap integral $Q$ is defined as $Q = \left\langle \rho_{01} \nabla_\perp^2 |E_{01}|^2 \right\rangle / \left\langle \rho_{01}^2 \right\rangle$, where $\rho_{01}$ and $E_{01}$ are the transverse field distributions of the R$_{01}$ mechanical mode and the LP$_{01}$ optical mode.

Now we consider the relative timing of the two solitons in the bound-state within one time-slot with respect to the index modulation induced by the acoustic wave, as shown in FIG. 2. The two solitons located at $t_1$ and $t_2$ is spaced by $\Delta t$, while one cycle of index modulation has a length of $T_a$ and is centered at the origin for simplicity. The timing of the double-soliton unit relative to the excited mechanical vibration in one time-slot can be expressed using

$$t_e = \frac{1}{\Omega_a} \tan^{-1}\left(-\frac{2\delta_\Omega}{\Gamma_B}\right), \qquad (2)$$

where $t_e$ is the equivalent driving center of the double-soliton unit, and the timing of the 1st and the 2nd solitons in this frame can be expressed as $t_1 = t_e + \Delta t/2$ and $t_2 = t_e - \Delta t/2$.

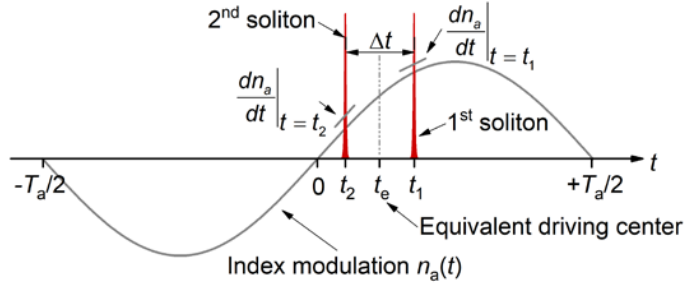

FIG. 2 One double-soliton unit of the ADS supramolecule and the optically-driven index modulation within one time-slot of the optomechanical lattice. In one time-slot, two solitons locate at $t_1$ and $t_2$, giving a soliton spacing of $\Delta t$ and an equivalent driving center of $t_e$. When propagating in the PCF core, the two solitons experience different refractive index slopes.

The optically-driven mechanical vibration modulates the effective refractive index of the LP$_{01}$ optical mode through the stress-optical effect, which can be expressed as

$$n_a(t) = \frac{\gamma_e \Theta}{2n_0 \rho_0} b(t) = C_a \frac{P_{av}}{\Omega_a \sqrt{4\delta_\Omega^2 + \Gamma_B^2}} \cos\left(\frac{\Omega_a \Delta t}{2}\right) \sin(\Omega_a t), \qquad (3)$$

where $C_a = \gamma_e^2 |Q|\Theta/(4\pi c n_0^2 A_{eff} \rho_0)$, $\rho_0$ is the material density of silica, and $\Theta$ is the overlap integral between the normalized intensity field of the LP$_{01}$ optical mode and the normalized density field of the R$_{01}$ mechanical mode, defined as $\Theta = \langle \rho_{01} |E_{01}|^2 \rangle / \langle \rho_{01}^2 \rangle$.

Back-action of the mechanical vibration results in a carrier frequency shift of the driving soliton, which is directly related to the refractive index slope that the soliton experiences as[22]:

$$\frac{\partial \omega_s}{\partial z} = -\frac{\omega_s}{c} \frac{\partial n_a}{\partial t}, \qquad (4)$$

where $\omega_s$ is the soliton carrier frequency. As shown in FIG. 2, in one mechanical cycle, the two solitons experience different index slopes, leading to a carrier frequency difference between the two solitons. Using Eq. (3) and Eq. (4), this frequency difference after a single-trip propagation in the PCF can be expressed as:

$$\Delta \omega_a = \delta \omega_{s2} - \delta \omega_{s1} = \frac{2 C_a \omega_s P_{av} L_{PCF}}{c} \frac{\delta_\Omega}{4\delta_\Omega^2 + \Gamma_B^2} \sin(\Omega_a \Delta t), \qquad (5)$$

where $\delta\omega_{s1}$ and $\delta\omega_{s2}$ are the carrier frequency shifts of the 1st and 2nd solitons due to the optomechanical effect in the PCF, and $L_{PCF}$ is the PCF length. When the lattice frequency of the soliton supramolecule is lower than the mechanical resonance frequency in the PCF ($\delta_\Omega < 0$), $\Delta\omega_a$ remains negative, which means that the optomechanical effect leads to a lower carrier frequency of the 2nd soliton than that of the 1st one. In a fiber cavity with anomalous average dispersion, solitons with lower frequency propagate at a lower group velocity, which delays the 2nd soliton relative to the 1st soliton during propagation, leading to an effective force of attraction between the two solitons.

### D. Force of repulsion through dispersive wave perturbations

A competing long-range interaction between solitons results from the dispersive waves that are shed from one soliton and extend in the time domain to the neighboring solitons and perturb them through cross-phase modulation[23]. Dispersive waves, which corresponds to Kelly sidebands in soliton fiber lasers, appear at discrete frequencies on the soliton spectrum (FIG. 3(a)). In the systems for long-distance soliton transmission and soliton fiber lasers, the existence of dispersive waves is a generic phenomenon that is due to periodic disturbances arising from the discrete dispersion, nonlinearity, gain and loss in these systems[24-26]. During propagation, optical solitons coherently transfer their energy to several discrete frequency components (sidebands). Due to gain filtering in the EDFA, those sidebands would always experience a net loss in the cavity. During one cavity round-trip, the nonlinear gain of the sidebands balances their net cavity loss due to energy conversion from solitons to sidebands, determining the steady-state intensities of the sidebands. In practice, due to the asymmetric gain profile in the EDFA and higher-order dispersion in the fiber cavity, these Kelly sidebands are usually asymmetrically distributed on the soliton spectrum[27]. The situation we consider in the main text is that the $m=-1$ order sideband has the dominant power as shown in FIG. 3(a) (Other possibilities are demonstrated experimentally in Section 2C below in which different relative powers of the Kelly sidebands are concerned).

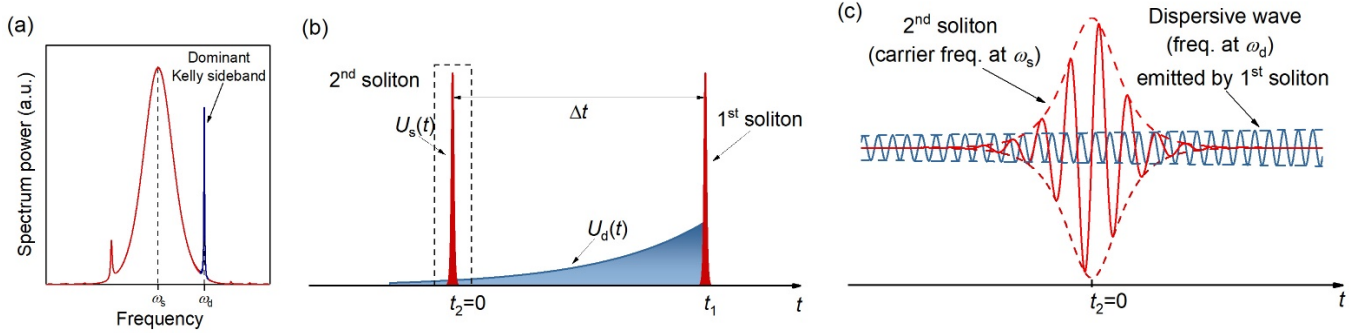

FIG. 3 Long-range soliton interactions due to dispersive wave perturbations. (**a**) Typical soliton spectrum with a dominant ($m=-1$ order) Kelly sideband at $\omega_d$ ($\omega_d > \omega_s$) highlighted in blue. (**b**) This dominant sideband corresponds, in the time domain, to dispersive waves shed from the 1st soliton, which has a faster group velocity than the soliton, and would eventually extends to the 2nd soliton. (**c**) Interactions between the 2nd soliton and the dispersive wave shed from the 1st soliton shift the carrier frequency of the 2nd soliton.

This dominant sideband corresponds, in the time domain, to a packet of dispersive waves shed from solitons in the laser cavity. Within one double-soliton unit, the dispersive wave shed from the 1st soliton (see FIG. 3(b)) travels faster than solitons due to its higher carrier frequency, and its exponentially-decaying envelope is due to the net cavity-loss that it experiences in the cavity. In this moving frame, the 2nd soliton can be expressed as:

$$U_s(t) = A_s \, \text{sech}[t/\tau_s] \exp[i\omega_s t - i\varphi_s(z)] = A_s u_s(t), \tag{6}$$

where $A_s$ is the soliton peak amplitude, $u_s(t)$ the normalized profile, $\tau_s$ the pulse width, and $\varphi_d$ the phase of the 2nd soliton varying with propagation length. For simplicity, we neglect the variation of soliton envelope within one round-trip, which means that the soliton duration (thus the bandwidth) and soliton peak amplitude are both considered as constant during propagation. In a reference frame moving with the 2nd soliton ($t_2 = 0$), the dispersive waves shed from the 1st soliton can be expressed as:

$$U_d(t) = A_d \exp[h(t-\Delta t)] \exp[i\omega_d(t-\Delta t) - i\varphi_d(z)] = A_s u_d, \tag{7}$$

where $A_d$ is the dispersive wave amplitude, $h$ is the decay rate, $\omega_d$ is the carrier frequency, and $\varphi_d(z)$ is the phase of the dispersive wave. Then $u_d$ is the dispersive wave waveform normalized to the soliton peak amplitude. Note that $\varphi_d(z)$ includes the propagation term of the dispersive wave, and varies as the wave propagates in the cavity. In practice, while $A_d$ and $\omega_d$ could be estimated using peak intensity and central frequency of the dominant sideband, $h$ could be calculated using the spectral bandwidth of the sideband.

According to soliton perturbation theory[23, 28], the carrier frequency of the 2nd soliton varies due to the dispersive wave shed from the 1st soliton (see FIG. 3(c)) following the expression

$$\frac{d\omega_s}{dz} = \frac{1}{\tau_s L_D} \text{Im}\left\{ \int_{-\infty}^{+\infty} u_s^*(t) \tanh(t/\tau_s) u_d(t) dt \right\}. \tag{8}$$

where $L_D$ is the dispersion length of the soliton. By substituting Eq. (6) and Eq. (7) into Eq. (8), we can obtain

$$\frac{d\omega_s}{dz} = \frac{A_d}{A_s \tau_s L_D} B(\Delta\omega) \exp(-h\Delta t) \cos[\Delta\varphi(z)], \tag{9a}$$

$$\begin{aligned} B(\Delta\omega) &= \int_{-\infty}^{+\infty} \text{sech}(t/\tau_s) \tanh(t/\tau_s) \sin(\Delta\omega t) dt \\ &= \pi \Delta\omega \tau_s \, \text{sech}(\pi \Delta\omega \tau_s / 2) \end{aligned}, \tag{9b}$$

where $\Delta\omega = \omega_d - \omega_s$ is the carrier frequency difference, and $\Delta\varphi(z) = \varphi_s(z) - \varphi_d(z)$ is the phase difference between the soliton and the dispersive wave. In practice, $B(\Delta\omega)$ is a constant close to 1, and the accumulated frequency shift of the 2nd soliton after each cavity round-trip (RT) can be expressed as:

$$\Delta\omega_{\mathrm{d}}^{\mathrm{RT}} = \frac{A_{\mathrm{d}}}{A_{\mathrm{s}}\tau_{\mathrm{s}}L_{\mathrm{D}}} B(\Delta\omega)\exp(-h\Delta t)\int_{z_0}^{z_0+L_{\mathrm{c}}} \cos[\Delta\varphi(z)]dz, \qquad (10)$$

where $z_0$ is the starting point for integral in the cavity, and $L_{\mathrm{c}}$ the cavity length. Note that in Eq. (10) we assume an invariant intensity of the dispersive wave during its propagation in the cavity. Otherwise we would have to include an intensity distribution $A_{\mathrm{d}}(z)$ within this integral.

The phase-matching condition for dispersive wave generations in a soliton laser[24, 25] requires that the accumulated phase difference between solitons and dispersive waves over one cavity round-trip should be integer multiples of $2\pi$. For the $m=-1$ order dispersive wave, this accumulated phase difference should be $-2\pi$. If we define the phase difference at an arbitrary position ($z_0$) in the cavity as $\Delta\varphi_0$, we obtain:

$$\Delta\varphi(z) = \int_{z_0}^{z}\left[\frac{1}{2}\beta_2(z)\Delta\omega^2 - \frac{1}{2}\gamma(z)|A_{\mathrm{s}}(z)|^2\right]dz + \Delta\varphi_0, \qquad (11a)$$

$$\Delta\varphi(z+L_{\mathrm{c}}) - \Delta\varphi(z) = -2\pi, \qquad (11b)$$

where $\beta_2(z)$ and $\gamma(z)$ are the dispersion and nonlinearity maps of the laser cavity. By substituting Eq. (11a) into Eq. (10), we can obtain:

$$\Delta\omega_{\mathrm{d}}^{\mathrm{RT}} = \frac{A_{\mathrm{d}}}{A_{\mathrm{s}}\tau_{\mathrm{s}}L_{\mathrm{D}}} B(\Delta\omega)\exp(-h\Delta t)\psi(\Delta\varphi_0), \qquad (12a)$$

$$\psi(\Delta\varphi_0) = \int_{z_0}^{z_0+L_{\mathrm{c}}} \cos\left\{\int_{z_0}^{z}\left[\frac{1}{2}\beta_2(z)\Delta\omega^2 - \frac{1}{2}\gamma(z)|A_{\mathrm{s}}(z)|^2\right]dz + \Delta\varphi_0\right\}dz. \qquad (12b)$$

Using Eq. (12b), it can be seen that when both the dispersion and nonlinearity distributions in the cavity are fixed, the integral $\psi$ only depends on the initial phase difference ($\Delta\varphi_0$), which determines the sign of the carrier frequency shift of the perturbed soliton. In order to form stable double-soliton units, an effective force of repulsion due to this dispersive wave perturbation is necessary to balance the force of attraction due to the optomechanical effect. According to Eqs. (12), this requires a special phase relationship between the dispersive wave shed from the 1st soliton and the 2nd soliton, which ensures a positive $\Delta\omega_{\mathrm{d}}^{\mathrm{RT}}$. Acting in concert with the anomalous average dispersion of the laser cavity, this positive $\Delta\omega_{\mathrm{d}}^{\mathrm{RT}}$ would lead to a faster group velocity of the 2nd soliton than that of the 1st one, resulting in an effective force of repulsion between the two solitons. The existence of this phase relationship has been verified in the experiments. As will be shown in FIG. 7(c) and FIG. 15(b) in the following section (and in Fig.3C in the ref [29]), we observed clear interferometric fringes on the laser spectra, which appeared only in the vicinity of the dominant sideband and quickly decayed outside this region, indicating the phase locking between the dispersive waves and the perturbed soliton.

We observed in the experiments that the interferometric fringes (and thus corresponding phase locking) were very robust over time. This stable phase locking might be attributed to the nonlinear coupling between the dispersive wave and the perturbed soliton. The concept of attractor in a nonlinear system might be introduced to investigate the robustness of this phase locking between the two optical waves. At present, we use this phase relation as a basic assumption of the theoretical derivation, and leave a full explanation of its origin as an open question.

### E. Balance between the two long-range forces: trapping potentials

The build-up of the long-range binding of the two solitons in one time-slot of the optomechanical lattice is based on the precise balance between the effective long-range forces of attraction and repulsion. Over each cavity round-trip, the overall carrier frequency shift of the 2nd soliton relative to the 1st soliton should equal to zero (i.e. $\Delta\omega_{\mathrm{a}}+\Delta\omega_{\mathrm{d}}^{\mathrm{RT}}=0$). Therefore the two solitons always have the same group velocity during propagation and travel together with an invariant soliton spacing. Using Eq. (5) and Eq. (12a), this balance can be expressed as:

$$-\frac{2C_{\mathrm{a}}\omega_{\mathrm{s}}P_{\mathrm{av}}L_{\mathrm{PCF}}}{c}\frac{\delta_\Omega}{4\delta_\Omega^2+\Gamma_{\mathrm{B}}^2}\sin(\Omega_{\mathrm{a}}\Delta t) = \frac{A_{\mathrm{d}}}{A_{\mathrm{s}}\tau_{\mathrm{s}}L_{\mathrm{D}}}A_{\mathrm{d}}B(\Delta\omega)\exp(-h\Delta t)\psi(\Delta\varphi_0). \qquad (13)$$

While the right side of the equation has a positive sign corresponding to a force of repulsion, a negative detuning of the lattice frequency from the mechanical resonance frequency ($\delta_\Omega<0$) ensures that the optomechanical effect in the PCF creates an effective force of attraction between the two solitons. The theoretical fitting curves in Fig.3D and Fig.S5C in ref [29] are both based on Eq.(13).

When the system parameters are fixed, it can be seen that the intensity of the force of attraction on the left side of Eq. (13) increases as the soliton spacing ($\Delta t$) increases, while the intensity of the force of repulsion on the right side of Eq. (13) decreases as the soliton spacing increases. Such line shapes are necessary to form a temporal trapping potential for the 2$^{nd}$ soliton (see FIG. 1,(d)–(f)), which makes the long-range soliton binding robust.

## F. Motion of soliton under the presence of trapping potential

Due to the presence of the trapping potential around the balanced positions emerged from the long-range bit-bit interactions, the timing-jitter of the supramolecular soliton stream was able to remain non-growing within time (as demonstrated in ref [29]). Such phenomenon differs itself from the conventional case in which the timing-jitter of a pulse stream would increase during propagation over optical fiber. We establish here a simple model that describes the motion of the soliton (timing jitter) under the presence of different types of "forces", including the long-range, non-covalent force, the damping-related force caused by gain filtering effect, and the noise-related force coming from the spontaneous emission of the EDFA ("heat bath"). Here we still use the double-soliton unit as our model. To simplify the notations, we locate the 1$^{st}$ soliton, being the reference soliton, at the origin (instead of $t_1$), while the 2$^{nd}$ soliton will located at the balanced position $t_0$. In order to simplify the mathematical expression of the following model, we will switch from the time-domain description to the spatial-domain description, while still using the moving frame, i.e. the spatial coordinate $z$ is related to the time coordinate $t$ via $z = z' - v_g t$, in which $z'$ is the actual spatial coordinate along the fiber. Then, the balanced pulse spacing in the time domain is directly related to a spatial pulse spacing of $z_0$ (see FIG. 4(a)) through $z_0 = -v_g t_0$.

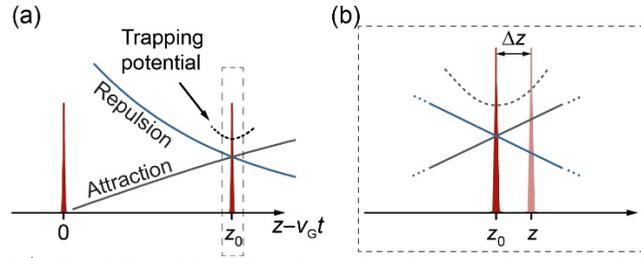

FIG. 4 (a) The trapping potential for the 2$^{nd}$ soliton is formed due to the balance of two long-range forces at $z_0$. (b) The slight deviation from the equilibrium position ($\Delta z$) would lead to a net restoring force.

The modulations on soliton carrier frequency due to optoacoustic effects and dispersive wave perturbations described above can be linearized in the close vicinity of the balanced position $z_0$ (see FIG. 4(b)) as below:

$$\left(\frac{d\Delta\omega}{dt}\right)_A = -K_A z \tag{14a}$$

$$\left(\frac{d\Delta\omega}{dt}\right)_D = C_0 - K_D z \tag{14b}$$

The two opposite frequency-shifts should precisely cancel each other at the equilibrium position $z_0 = C_0/(K_A + K_D)$. In the vicinity of this equilibrium position, if the soliton has a slight spatial deviation $\Delta z = z - z_0$, the soliton would have a net frequency shift during propagation, i.e.

$$\frac{d\Delta\omega}{dt} = -K\Delta z \tag{15}$$

where $K = K_A + K_D$. The spatial deviation $\Delta z$ is related to the group velocity difference $\Delta v_g$ between the two solitons through

$$\frac{d\Delta z}{dt} = \Delta v_g \tag{16}$$

and $\Delta v_g$ is related to $\Delta\omega$ through the cavity group velocity dispersion, i.e.

$$\Delta v_g = B\Delta\omega, \text{ where } B = -\beta_2 v_g^2 \tag{17}$$

Then we substitute Eq.(17) into Eq.(15) and we can obtain

$$\frac{d\Delta v_g}{dt} = -KB\Delta z \qquad (18)$$

In addition to these long-range forces described above, the solitons also experiences the gain filtering effect in the EDFA. Such filtering effect actually provides a "damping force" upon the solitons, which can be described as[21, 30, 31]:

$$\frac{d\Delta\omega}{dt} = -\Gamma\Delta\omega \qquad (19)$$

Using Eq. (16), we can readily obtain:

$$\frac{d\Delta v_g}{dt} = -\Gamma\Delta v_g \qquad (20)$$

Then a random term $S_{\Delta\omega}$ should be introduced to describe the white noise source, which results from the amplified spontaneous emission (ASE) of the EDFA and can be written as[31]:

$$\begin{cases} \langle S_{\Delta\omega} \rangle = 0 \\ \langle S_{\Delta\omega}(t) S_{\Delta\omega}(t') \rangle = N_{\Delta\omega}\delta(t-t') \end{cases} \qquad (21)$$

where the bracket $\langle ... \rangle$ stands for the time-domain integral, $N_{\Delta\omega}$ describes the noise level, and the delta function simply indicates that noise at different time is uncorrelated[31].

Now, we have the following coupled equations that can describe the motion of the second soliton in the vicinity of the equilibrium position:

$$\begin{cases} \dfrac{d\Delta z}{dt} = \Delta v_g \\ \dfrac{d\Delta v_g}{dt} = -KB\Delta z - \Gamma\Delta v_g + S_{\Delta\omega} \end{cases} \qquad (22)$$

For simplicity, we replace $(\Delta z, \Delta v_g)$ with $(z, v)$ and then we can actually have the well-known Langevin equations in the presence of a harmonic potential (due to the "spring" effect). The mean square deviation of $z$ can then be obtained to be:

$$\langle z(t) - z_0 \rangle^2 = \frac{N_{\Delta\omega}}{2\Gamma KB} \qquad (23)$$

As we can reveal from Eq.(23). The presence of the harmonic potential causes that the mean square deviation of the relative pulse position (timing jitter) no longer grows with time, as depicted in FIG. 5(b). The experimental results demonstrated in ref [29] have confirmed this remarkable feature of the supramolecular soliton stream.

In contrast, when the spring effect (harmonic potential) is absent (i.e. the term $-KB\Delta z$ is absent), Eqs.(22) would degrade into the conventional Langevin equations that describe the Brownian motion of a particle in a heat bath (i.e. liquid). The mean square deviation of the particle position $z(t)$ can be expressed using the well-known diffusion equation as:

$$\langle z(t) - z_0 \rangle^2 = \frac{N_{\Delta\omega}}{\Gamma^2} t \qquad (24)$$

In this case the mean square deviation of the relative pulse position would increase linearly with time as depicted in FIG. 5(a), corresponding to the well-known Gordon-Haus jitter in soliton telecommunications systems[30].

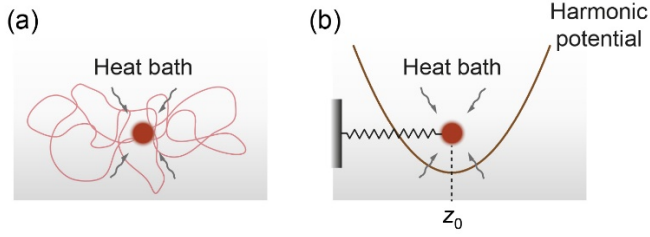

FIG. 5. (**a**) Random walk of a particle in the heat bath with a mean square deviation growing linearly with time. (**b**) A particle trapped within a harmonic potential, whose displacement has a constant mean square deviation despite continuous perturbations of the heat bath.

## III. Experiments

### A. Homogeneous supramolecular assembly of optical solitons

Experimental generation of soliton supramolecular assembly in the optoacoustic mode-locking fiber laser depends on both the cavity design (fiber sections with different parameters and lumped elements) and the self-starting process which relied on the careful adjustment of the working point of the laser cavity[21]. In our experiments, we were able to partially control the pattern of the soliton supramolecule generated in our fiber laser through varying the working point, tuning the laser parameters (pump power, etc.) and/or changing the cavity design. The simplest cases which can repeatedly generate are the homogeneous soliton supramolecules in which all the time-slots within one round-trip are filled with identical long-range bound-states of multiple solitons. The case of the simplest all-double-soliton (ADS) has been demonstrated in ref [29]. By slightly increasing the pump power and adjusting the working power, supramolecular soliton stream with homogeneous all-triple-soliton (ATS) units can also be obtained in our experiments. We plot the time-domain trace of a typical ATS supramolecule over one cavity round-trip in FIG. 6(a) such that the consecutive time-slots are plotted in parallel, in order to appreciate the fact that all the triple-soliton units share the same internal spacings as well as relative timing within their time-slots. The persistence-mode plot over one time-slot span is shown in FIG. 6(b), which shows that homogeneous internal spacing in all the time-slots. The internal spacing between the $1^{st}$ and $3^{rd}$ soliton is 147 ps, while the internal spacing between consecutive solitons are 76 ps ($1^{st} - 2^{nd}$) and 71 ps ($2^{nd} - 3^{rd}$). The FFT power spectrum of the ATS supramolecule stream is shown in FIG. 6(c), in which the modulated envelope matches the internal soliton spacing of the triple-soliton units.

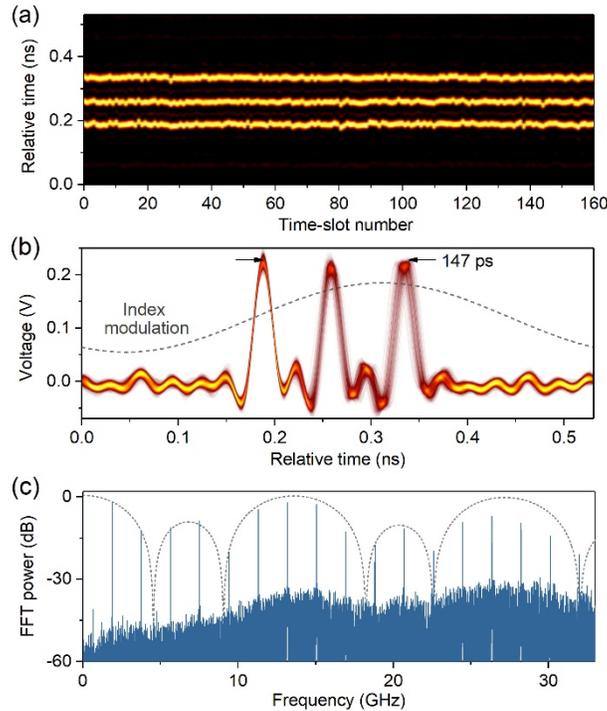

**FIG. 6.** (**a**) One round-trip of the time-domain trace of the ATS supramolecule, with all the 160 time-slots plotted in parallel. (**B**) The same time trace plotted under the persistence mode over one time-slot span. (**C**) The FFT power of the ATS supramolecule soliton stream.

The autocorrelation trace of the ATS supramolecule stream is shown in FIG. 7(a), showing an individual soliton duration of 690 fs. The optical spectrum of the ATS supramolecular stream is shown in FIG. 7(b), where the characteristic spectrum fringe also appears in the vicinity of the dominant sideband (see FIG. 7(c)), and the fringe period matched the internal spacing between the solitons within the bound-state, similar to the case in the ADS-case in ref [29]. Note that the fringe has a slightly weak contrast, due to the fact of the slightly different internal spacing between the three solitons, which smears out each other in the optical spectrum. The sideband on the other side (m=+1 order), as expected, still exhibits no fringe. These experimental results, together with these from the ADS-case, confirm the fact that the repulsive force are indeed related to the dominant Kelly sideband of the soliton, a fixed phase-relation was required under the balanced state, as predicted in the theory section above.

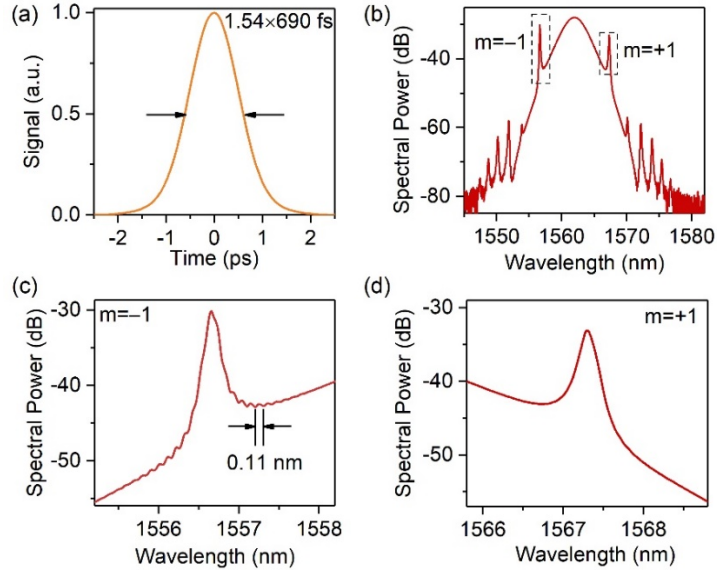

FIG. 7. (**a**) The autocorrelation trace and (**b**) the optical spectrum of the ATS supramolecular stream. (**c**) The zoom-in of the dominant (m=−1-order) sideband of the spectrum in (b). (**d**) The zoom-in of the sideband at m=+1 order of the spectrum in (b).

Generation of quadrupole-soliton units in such a supramolecular soliton stream usually demands a higher pump power. The time-domain trace of the supramolecular stream that contains quadrupole-soliton units and triple-soliton units are partially shown in FIG. 8(a), and the same pattern can also be preserved after each cavity round-trip (FIG. 8(b)). This pattern could also be stably preserved over 20 minutes, even though the quadrupole-soliton units were slightly unstable compared to the double- and triple-soliton units. This is perhaps due to a weaker trapping potential that has becomes more vulnerable to noise perturbations. It might also be caused by the large spacing between the 1$^{st}$ and 4$^{th}$ soliton within one unit, which could result in that perturbations of dispersive waves emitted from solitons in neighbouring time-slots might become non-trivial.

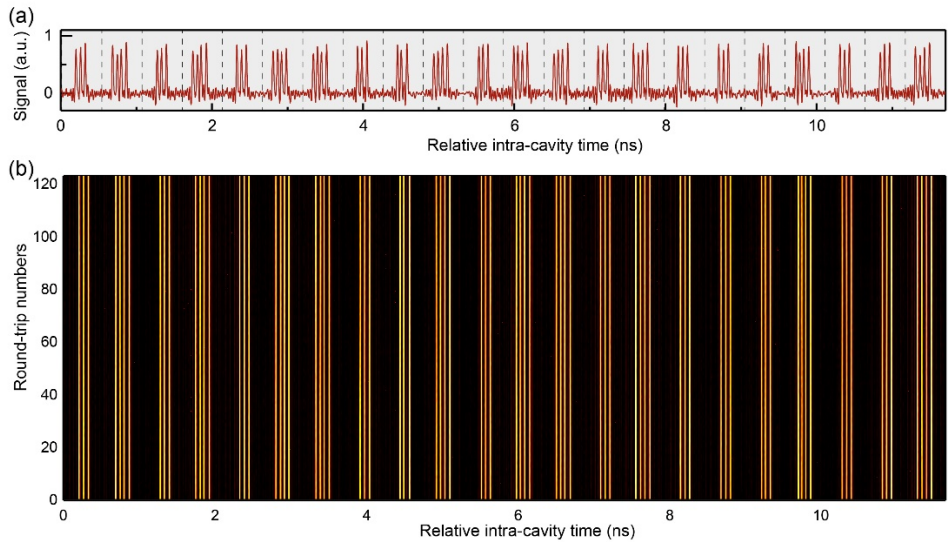

FIG. 8. (**a**) The time-domain trace of the soliton supramolecule that contains only triple- and quadrupole-soliton units (only 23 out of 160 time-slots are plotted) (**b**) The round-trip plot of the same time-domain trace in (a) over 123 round-trips (10 μs).

## C. Directions of the repulsive forces

This section aims to elucidate the full picture of the possible time-domain relation of the multiple solitons bound within one time-slot of the optomechanical lattice. Previously we only illustrate one possibility of the asymmetric Kelly-sideband distributions, in which only the m=−1-order sideband is the dominant sideband. The asymmetry of the laser spectrum could be induced by the asymmetric gain profile in the EDFA[32] as well as the higher-order dispersion of the laser cavity[27]. When we assume that the two Kelly-sidebands symmetrically located on the soliton spectrum (the absolute values of their deviations from the soliton central frequency are equal), asymmetric gain profile in the EDFA would impose different gain coefficients to the two first-order ($m=\pm1$) Kelly-sidebands, leading to a difference between their stationary intensities. On the other hand, higher-order dispersion in the cavity would lead to an asymmetric distribution of the two sidebands on the soliton spectrum, resulting in a difference in energy coupling from solitons[27, 33]. In the experiments we could adjust the relative intensities of the two first-order sidebands by carefully adjusting the gain value in the EDFA and the high-order dispersion in the cavity.

Variations of relative intensities of the two sidebands could change the configuration of the repulsive force which is induced by dispersive wave perturbations. By illustrating the hybrid soliton supramolecules (with one soliton or two long-range bound-state solitons trapped in the time-slots of the lattice) using the "time-slot plot" similar to the plot in FIG. 6(a), we can observe more information of the effective long-range force of repulsion between the two solitons due to dispersive wave perturbations. Three different cases of the relative sideband intensities has been achieved in experiments and the results are demonstrated in FIG. 9 below.

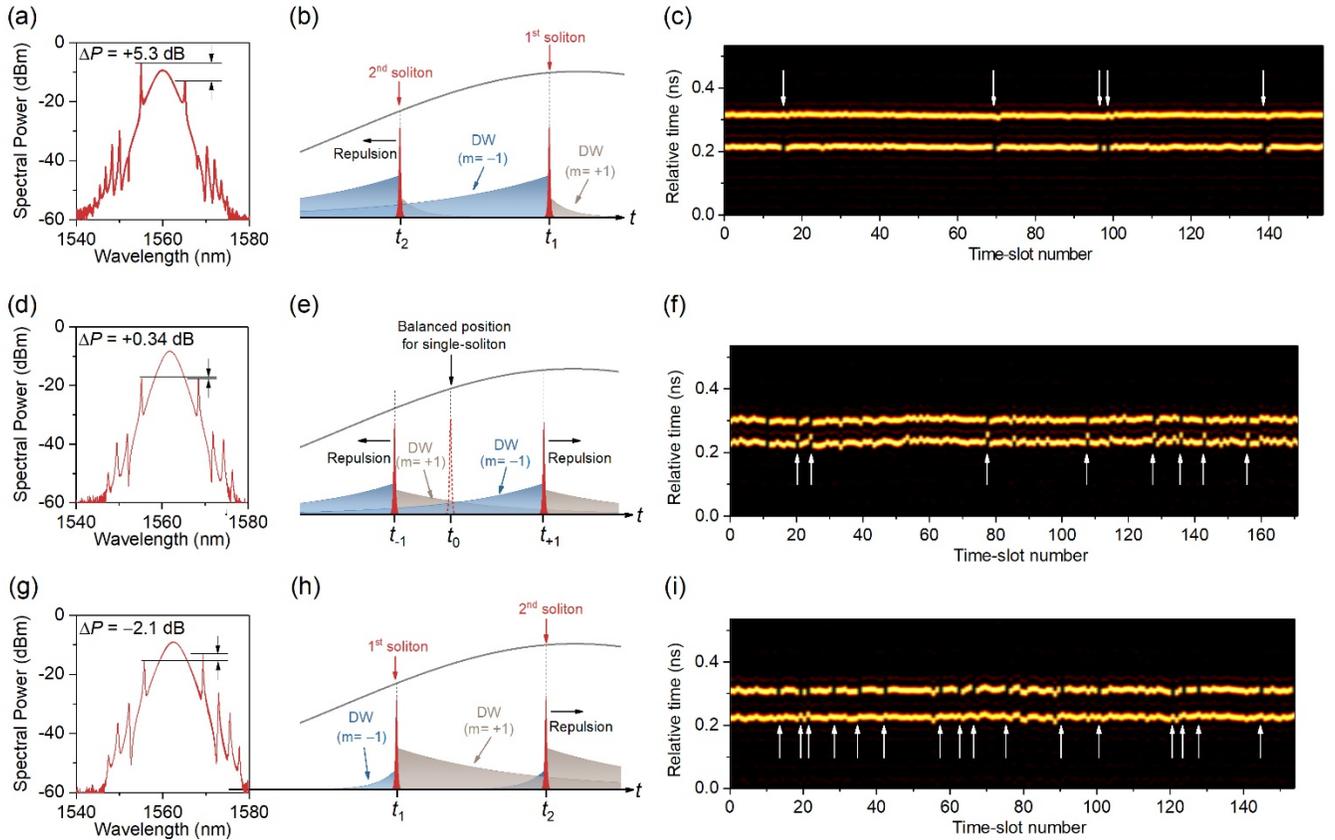

FIG. 9. Three cases of the force of repulsion due to dispersive wave perturbations. (**a**) The case that the $m=-1$ order sideband dominates. (**b**) The dispersive waves shed by the soliton are located later in time (set as the 1$^{st}$ soliton) perturbs the soliton located earlier in time, leading to an effective force of repulsion (**c**) The hybrid soliton supramolecule with several single-soliton units. The positions of those single-soliton units are marked by the white arrows (the upper row of pulses) (**d**)–(**f**) The two first-order ($m=\pm1$) sidebands have comparable intensities in the second case. Both of these sidebands (dispersive waves) will contribute a force of repulsion between the two solitons, and push the two solitons in the double-soliton unit away from a balanced position. (**g**)–(**i**) When the $m=+1$ sideband dominates, the 1$^{st}$ soliton is earlier in time. The dispersive wave shed from this soliton travels slower than the soliton, and perturbs the 2$^{nd}$ soliton that is located later in time. We then find that the solitons in single-soliton units that are locates at the same relative position as the 1$^{st}$ soliton in the double-soliton unit (the lower row of pulses in (i)).

In the first case, as shown in FIG. 9(a), when the higher-frequency ($m=-1$ order) sideband dominates with an intensity difference $\Delta P=+5.3$ dB, we set the soliton located later in time as the reference (1$^{st}$) soliton (see FIG. 9(b)). This is because that the force of repulsion in the double-soliton unit is merely induced by perturbation of the dispersive wave shed from the 1$^{st}$ soliton, and the $m=+1$ order dispersive wave shed from the 2$^{nd}$ soliton is much weaker. When a single soliton is trapped in one time-slot of the lattice, it is located in the same position as that of the 1st soliton in the double-soliton unit, since neither of them experiences significant perturbations from dispersive waves. As shown in FIG. 9(c), in the experiments we recorded a hybrid soliton supramolecule including several time-slots with only one soliton trapped. In those single-soliton units (marked by white arrows), the solitons always locate at the "upper row" which corresponds to the later (the 1st soliton) position in FIG. 9(b). In the second case, as shown in FIG. 9(d), comparable intensities of the two ($m=\pm 1$-orders) sidebands ($\Delta P=+0.34$ dB) lead to two comparable perturbations of dispersive waves on both of the two solitons (FIG. 9(e)), which push both of the two solitons away from the balanced position as shown in FIG. 9(f). The third case is that the lower-frequency ($m=+1$ order) sideband dominates the force of repulsion with $\Delta P=-2.1$ dB (see FIG. 9(g)). In contrast to the first case shown in FIG. 9(a)–(c), we use the soliton located earlier in time as the reference (1st) soliton in this case. The $m=+1$ order dispersive wave shed from the 1st soliton travels slower and perturbs the 2nd soliton located later in time (FIG. 9(h)). As a consequence, in single-soliton units the solitons always locate in the lower row (see FIG. 9(i)) which corresponds to the earlier soliton position in FIG. 9(h).

We emphasize here that in most of our experiments the $m=-1$ order sideband had the highest intensity and dominated the force of repulsion as shown in FIG. 9(a). In the experiments, we did achieve almost uniform two ($m=\pm 1$-order) sidebands as shown in FIG. 9(d) and stronger +1-order sideband as shown in FIG. 9(g) through significantly increasing the cavity loss and simultaneously varying the cavity higher-order dispersion by inserting some dispersion compensation fiber into the laser cavity. The experimental results shown in FIG. 9 did not, however, alter the physical picture of the soliton supramolecules: two long-range, non-covalent soliton interactions work together to make a large number of optical solitons assemble into supramolecular structures. Moreover, these results exhibit that the supramolecular soliton assembly could be formed within a broad range of system parameters.

### D. Internal spacing tuning by varying the long-range forces

In the ref [29], we have demonstrate that by tuning the intensities of dispersive wave and acoustic wave, the inter-soliton spacing of the double-soliton bound-state in the ADS-case can be continuously tuned over a wide range. The dispersive wave was tuned through the lumped attenuation inserted in the cavity, which introduced tunable gain bandwidth filtering effect. The acoustic wave was tuned by simply changing the cavity length, therefore the repetition rate of the optomechanical lattice can be tuned within the optoacoustic gain band of the PCF.

Similar tuning experiment has also been done for the all-triple-soliton (ATS) supramolecule composed of triple-soliton units within all the time-slots of the lattice, as depicted in FIG. 10(a). In the experiments we observed that as we gradually varied the intensity of the $m=-1$ order dispersive wave (see FIG. 10(b)), the whole duration of the triple-soliton unit ($\Delta t_{13}$ in FIG. 10(c)) could be tuned within a range of 102 ps to 187 ps and the spacing between the 1st and the 2nd soliton ($\Delta t_{12}$ in xx(c)) could be tuned within a range of 50 ps to 95 ps. Note that the spacing between the consecutive solitons are not necessarily the same in the triple-soliton unit, and during the tuning experiments, the two spacings actually follows slightly different curves as we can see in FIG. 10(c).

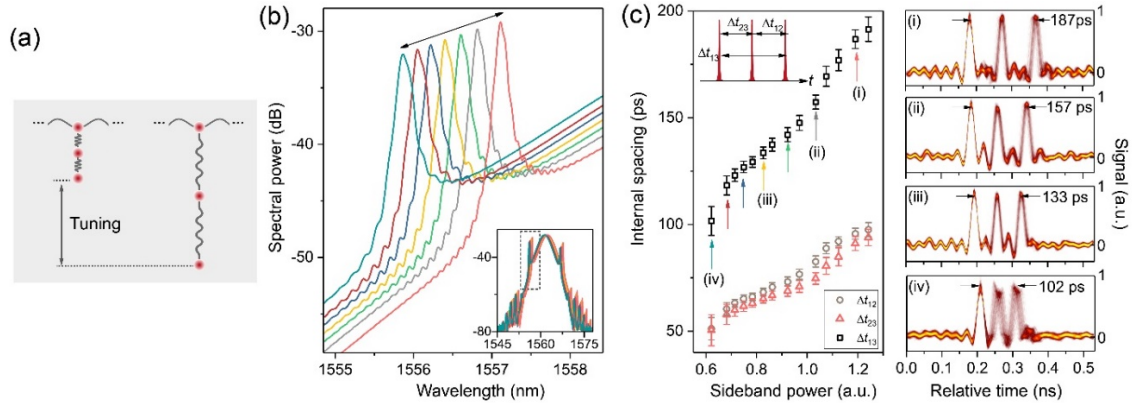

FIG. 10. Soliton spacing tuning of the all-triple-soliton supramolecule. (**a**) Tailoring of the intensity of the $m=-1$ order dispersive wave without changing spectral widths of the solitons (see the inset). (**b**) The soliton spacing increases as the dispersive wave grows. $\Delta t_{12}$, $\Delta t_{23}$ and $\Delta t_{13}$ are defined in the inset. Four examples are plotted in left columns (i) – (iv).

### E. Time-stretched dispersive Fourier transform measurement

The fundamental element of the soliton supramolecule, as demonstrated in ref [29], can be single solitons or phase-locked soliton pairs. Within a single time-slot, a single soliton and a phase-locked soliton pair (with internal spacing of ~ 4 ps) can form a long-range bound-state (with internal spacing of ~ 70 ps). Such compound structure can be detected using the well-known time-stretched dispersive Fourier transform[5, 34] (TS-DFT). Under the DFT measurement, the phase-locked soliton pair would be stretched to a broad pulse with strong fringes upon it, while single solitons would only be stretched to a relatively smooth profile. Beyond this example, we have also realized that the two soliton-pairs can further form a long-range bound-state. The experimental record of a soliton supramolecule that contains many of such units is shown in FIG. 11(a) (only part of the structure is shown). The time slots occupied by the long-range bound-state formed by two phase-locked soliton pairs are marked by the arrow. Such state is stably maintained after each round-trip, as shown in the continuous recording of the same structure over hundreds of round-trip in FIG. 11 (b). Then we sent this soliton stream into a long dispersive fiber (2-km SMF-28) and the resultant DFT signal is shown in FIG. 11 (c) and (d). We can readily realize that all the single-peaked pulses recorded in time-domain trace (in FIG. 11(a)) were actually phase-locked soliton pairs with the same internal spacing, due to the identical DFT signal. Specifically, in the time-slots with two soliton-pairs, the DFT signal features a weaker fringe contrast, due to the overlapping and thus smearing of the DFT signal of two soliton-pairs bound at long range.

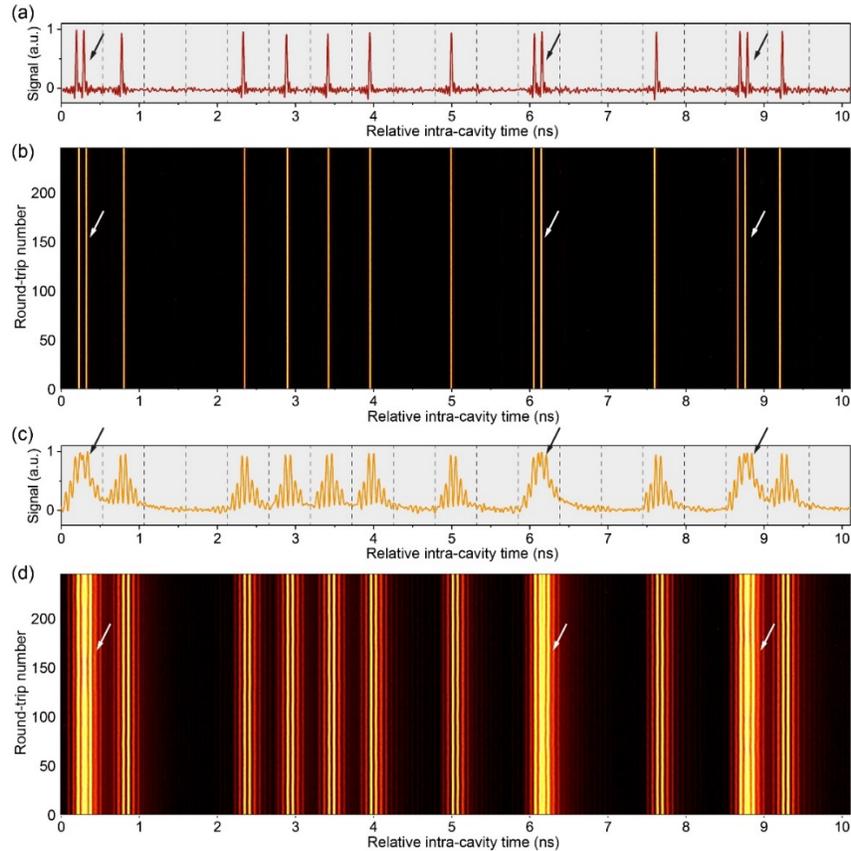

FIG. 11. (**a**) The soliton supramolecular stream that consists of "double-soliton"-units that are formed by two phase-locked soliton pairs (as marked by the black arrows). Only 19 out of 154 time slots are plotted. (**b**) The round-trip plot of the same time domain trace in (A) over 246 round-trips. (**c**) The DFT signal of the time-domain trace after 2-km SMF-28, featuring strong interferometric fringes. (**d**) The round-trip plot of the recorded DFT signal over 246 round-trips.

As an interesting comparison, we also record the DFT signal of a typical soliton supramolecular stream in which elementary components are only single solitons. The results are shown in FIG. 12, from which we can reveal that, in the time slots that accommodate double- and triple-soliton units (FIG. 12, (a) and (b)), the DFT signal is always simple, in form of incoherent adding of multiple overlapping DFT signals as shown in FIG. 12(c) and (d) without any obvious fringe. We would like to remark here one of the most important differences between the supramolecular soliton streams and the conventional soliton pairs or molecules: the solitons involving in long-range, non-covalent interactions are not necessarily phase-locked. In contrast, the formation of stable soliton pairs or molecules rely on direct, covalent interactions of the solitons at their pulse tails, resulting in robust phase locking between the solitons.

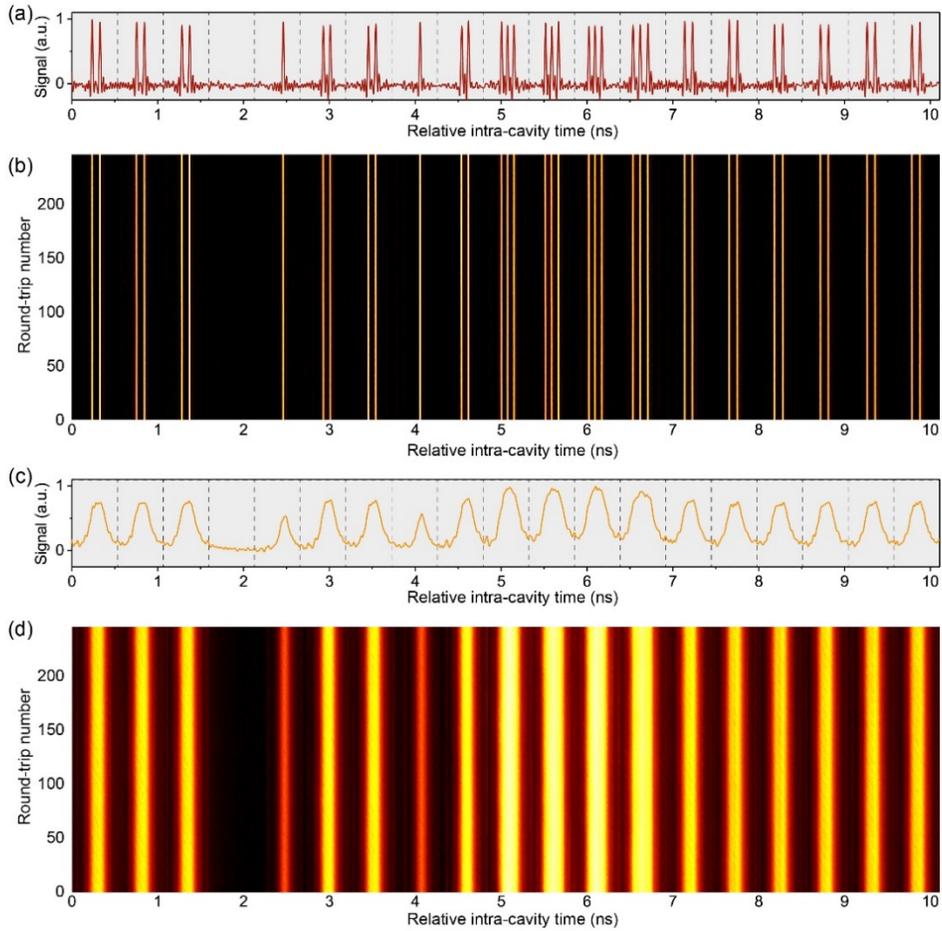

FIG. 12. (**a**) A typical supramolecular soliton stream that consists of multi-soliton units, while these units are configured only by individual solitons (no soliton-pairs) as fundamental elements. Only 19 out of 154 time-slots are shown due to limited figure size. (**b**) The persistence-mode plot of the same stream over 246 round-trips. (**c**) The corresponding DFT signal of the soliton stream in (**a**) using 2-km SMF-28. (**d**) The persistence-mode plot of the same DFT signal over 246 round-trips.

**F. Self-adjustment during adding and removing solitons**

The detailed pattern of the soliton supramolecules can be changed through strong perturbations, while the structure before and after the perturbation are both stable. In ref [29] we have demonstrated the experimental recording of the processes during adding and removing individual solitons in such supramolecular structure. However, in all those examples, the newly appeared solitons directly reached the right positions in the trapping potential when they were generated from noise background. This is however only a special case. In the experiments we have also observed that, by abruptly increasing the pump power (FIG. 13(a)) the newly-added solitons could initially appear away from the "correct" positions (as marked by the white arrows in FIG. 13(b)). Nevertheless, due to the presence of the long-range forces as described above, these newly-generated solitons would be quickly dragged backed to the "correct" (balanced) positions, and the supramolecular could still stably switch to a new pattern.

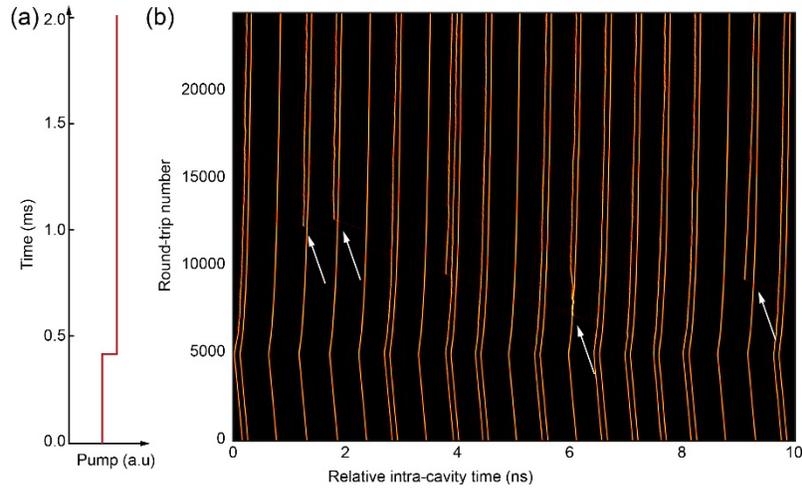

FIG. 13. The self-adjustment of internal spacing of multi-soliton units during the process of adding new solitons. (**a**) An abrupt increase of pump power by ~15% with a 5-ns rising edge. (**b**) Some new solitons that appeared from the background were initially far away from the balanced positions (or trapping potential centers), as marked by the white arrow. However, over a few thousands of round-trips these new solitons would still be shifted to the balanced positions due to the presence of the long-range forces.

Similar phenomenon has also be observed in some soliton-removing experiments. One example is shown in FIG. 14 in which the pump power was suddenly decreased by ~15%, and some existing solitons were removed from their time-slots. Interestingly, the quadruple-soliton units as marked by the white arrow in FIG. 14(b) lost its $3^{rd}$-soliton. After this abrupt change, the $4^{th}$-soliton just moved to the right until it reached the balanced position of the original $3^{rd}$-soliton. Then the quadruple-soliton unit then stably evolved into a triple-soliton unit.

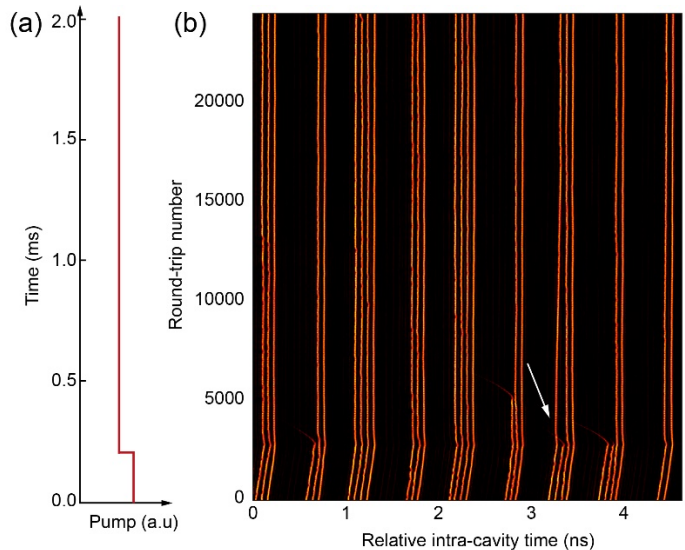

FIG. 14. The self-adjustment of internal spacing of multi-soliton units during the process of removing solitons. (**a**) An abrupt decrease of pump power by ~15% with a 5-ns trailing edge. (**b**) Some existing solitons are removed from their time-slots. The quadruple-soliton unit marked by the white arrow has lost its $3^{rd}$-soliton, then the $4^{th}$-soliton then drifts and re-fill the empty trapping potential and become the new $3^{rd}$-soliton.

At last we present one more case of dynamical process of the supramolecular soliton structure. In this case, the pump power was modulated such that it has an abrupt "dip" of ~ 50 μs, as shown in FIG. 15(a). Then we observed that the $2^{nd}$-soliton of a double-soliton unit has escaped from its original unit, and then re-appeared in another double-soliton unit, and then upgraded this unit into a triple-soliton unit (as highlighted by the white box in FIG. 15 (b)). We can see that the soliton intensity slightly decreased during the inter-unit transition, however it was not completely diminished and has actually re-gained its intensity later due to the recovery of the pump power. Longer transition distance have also be observed in our experiments, in which the escaped solitons could drift over a few units (~0.5 m in terms of relative distance) and joined in another unit (see FIG. 15 (c)). These experimental results provide one more possibility for controlling the fine structure of the supramolecular soliton stream besides simply adding or removing solitons.

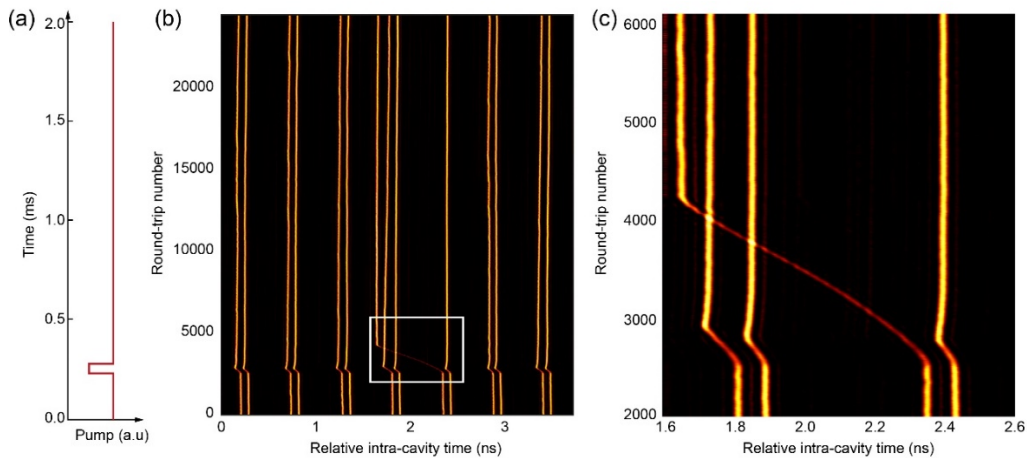

FIG. 15. Transition of individual soliton between different units. (**a**) An abrupt "dip" of the pump power by ~25% over 5 μs duration. (**b**) The 2$^{nd}$-soliton of one double-soliton unit has escaped from its original unit and was later trapped in another double-soliton unit, and then upgraded this unit into a triple-soliton unit.

## IV. Conclusion

By carefully tailoring the long-range interactions between the optical solitons in an optoacoustic mode-locking fiber laser, a large number of individual solitons can mutually configure a macroscopic supramolecular structure, simultaneously featuring dynamic stability, elementary diversity, and structural flexibility. We developed a theoretical model of such non-covalent bonds between optical solitons, and demonstrated a series of experimental results to support this model.